\documentclass[11pt]{article}
\usepackage{fullpage,graphicx}

\def\be{\begin{equation}}
\def\ee{\end{equation}}

\begin{document}
\title{General rogue waves in the focusing and defocusing Ablowitz-Ladik equations}
\author{Yasuhiro Ohta$^{1}$ and Jianke Yang$^{2}$ \\
{\small\it $^1$Department of Mathematics, Kobe University, Rokko,
Kobe 657-8501, Japan} \\ {\small\it $^2$Department of Mathematics
and Statistics, University of Vermont, Burlington, VT $05401$,
U.S.A}}
\date{}
\maketitle

\noindent\textbf{Abstract}\hspace{0.2cm}  General rogue waves in the
focusing and defocusing Ablowitz-Ladik equations are derived by the
bilinear method. In the focusing case, it is shown that rogue waves
are always bounded. In addition, fundamental rogue waves reach peak
amplitudes which are at least three times that of the constant
background, and higher-order rogue waves can exhibit patterns such
as triads and circular arrays with different individual peaks. In
the defocusing case, it is shown that rogue waves also exist. In
addition, these waves can blow up to infinity in finite time.

\section{Introduction}
Rogue waves are large and spontaneous nonlinear waves which ``come
from nowhere and disappear with no trace". These waves have drawn a
lot of attention in the nonlinear wave community recently since they
are linked to damaging freak waves in the ocean and transient
high-intensity optical waves in fibers
\cite{rogue_water,rogue_nature1}. Explicit expressions of rogue
waves have been derived for a large number of nonlinear integrable
systems. Examples include the nonlinear Schr\"odinger equation,
\cite{Peregrine,Akhmediev_PRE,Rogue_higher_order,Rogue_higher_order2,Rogue_Gaillard,Rogue_triplet,Rogue_circular,Liu_qingping,
OY,Matveev_Kyoto}, the derivative nonlinear Schr\"odinger equation
\cite{rogue_DNLS1, rogue_DNLS2}, the three-wave interaction equation
\cite{three_wave}, the Davey-Stewartson equations
\cite{OY_DSI,OY_DSII}, and many others
\cite{Ankiewicz2010,rogue_Hirota,rogue_Hirota2,Manakov1,Manakov2,Manakov3,Fokas_Lenells,NLS_MB,KE}.
Experimental observations of rogue waves have also been reported in
optical fibers and water tanks
\cite{Rogue_nature2,NLS_rogue_water1,NLS_rogue_water2}.

Almost all rogue wave solutions reported so far are for continuous
wave equations. Discrete wave equations, on the other hand, are also
important since they can model various physical systems such as wave
dynamics in optical lattices. Then an interesting question is rogue
wave behaviors in discrete systems. For the focusing Ablowitz-Ladik
equation and discrete Hirota equation, fundamental rogue waves and
certain special second-order rogue waves were derived in
\cite{Ankiewicz2010}. It was shown that these rogue waves can reach
higher peak amplitudes compared to their continuous counterparts.

In this paper, we derive general arbitrary-order rogue waves in the
focusing and defocusing Ablowitz-Ladik equations by the bilinear
method. These solutions are presented through determinants, and they
contain $2N+1$ non-reducible free real parameters, where $N$ is the
order of rogue waves. In the focusing case, we show that rogue waves
are always bounded. In addition, fundamental rogue waves reach peak
amplitudes which are at least three times that of the constant
background, and higher-order rogue waves can exhibit patterns such
as triangular and circular arrays with different individual peaks.
In the defocusing case, we show that rogue waves still appear, which
is surprising. In this case, we find that rogue waves of every order
can blow up to infinity in finite time, even though non-blowup rogue
waves also exist.

\section{General rogue-wave solutions}

The Ablowitz-Ladik (AL) equation has two types, the focusing and
defocusing ones. The focusing AL equation can be written as
\cite{ALpaper,ALpaper2}
\begin{equation} \label{e:ALf}
i\frac{d}{dt}u_n=(1+|u_n|^2)(u_{n+1}+u_{n-1}),
\end{equation}
and the defocusing AL equation is
\begin{equation} \label{e:ALd}
i\frac{d}{dt}u_n=(1-|u_n|^2)(u_{n+1}+u_{n-1}).
\end{equation}

Regarding rogue waves in these AL equations, we have the following
theorems.

\noindent\textbf{Theorem 1}\hspace{0.2cm} General $N$-th order rogue
waves in the Ablowitz-Ladik equations (\ref{e:ALf})-(\ref{e:ALd})
are given by
\begin{equation}   \label{f:general}
u_n(t)=\frac{\rho}{\sqrt{1-\rho^2}}\:  \frac{g_n}{f_n}e^{i(\theta n-\omega t)},
\end{equation}
where $\rho$ and $\theta$ are free real constants,
$\omega=2\cos\theta/(1-\rho^2)$,
\[
f_n=\tau_n(0), \quad g_n=\tau_n(1)/(1+\rho)^{2N},
\]
\[
\tau_n(k)=\left.\det_{1\le i,j\le N}\left(m_{2i-1,2j-1}^{(n)}(k)\right)
\right|_{p=q=1+\rho},
\]
\[
m_{ij}^{(n)}(k)=A_iB_jm^{(n)}(k),
\]
\[ m^{(n)}(k)=\frac{1}{pq-1+\rho^2}(pq)^n
\left(\frac{1-\rho^2-q}{1-1/p}\right)^k
e^{i\left(\frac{1}{pq}-\frac{1}{1-\rho^2}\right)\left(q\hspace{0.03cm} e^{i\theta}-p \hspace{0.03cm} e^{-i\theta}\right)t},
\]
\[
A_i=\sum_{\nu=0}^i\frac{a_{\nu}}{(i-\nu)!}[(p-1)\partial_p]^{i-\nu},
\]
\[
B_j=\sum_{\mu=0}^j\frac{\bar a_{\mu}}{(j-\mu)!}[(q-1)\partial_q]^{j-\mu},
\]
$a_{\nu}$ are complex constants, overbar $\bar{}\ $ represents
complex conjugation, and
\begin{equation}  \label{e:aeven}
a_0=1, \quad a_2 =a_4=\cdots=a_{\mbox{\small even}} =0.
\end{equation}
When $|\rho|<1$, these rogue waves satisfy the focusing AL equation
(\ref{e:ALf}); and when $|\rho|>1$, they satisfy the defocusing AL
equation (\ref{e:ALd}).

The above expression (\ref{f:general}) for rogue waves involves
differential operators $A_i$ and $B_j$. A more explicit and purely
algebraic expression for these rogue waves (without the use of such
differential operators) is presented in the following theorem.

\noindent{\textbf{Theorem 2}} \hspace{0.1cm} General $N$-th order
rogue waves (\ref{f:general}) for AL equations can be rewritten as
\begin{equation} \label{f:theorem2}
u_n(t)=(-1)^N \frac{\rho}{\sqrt{1-\rho^2}}\:
\frac{\sigma_n(1,0)}{\sigma_n(0,0)} \hspace{0.15cm}  e^{i(\theta n-\omega t)},
\end{equation}
where $\rho$, $\theta$ and $\omega$ are the same as those in Theorem
1,
\begin{equation}  \label{def:sigman}
\sigma_n(k,l)=\det_{1\le i,j\le N}
\left(\widetilde m_{2i-1,2j-1}^{(n)}(k,l)\right),
\end{equation}
\[
\widetilde m_{ij}^{(n)}(k,l)=\sum_{\nu=0}^{\min(i,j)}
\left(\frac{1-\rho}{1+\rho}\right)^\nu
\Phi_{i\nu}^{(n)}(k,l)\Psi_{j\nu}^{(n)}(k,l),
\]
\[
\Phi_{i\nu}^{(n)}(k,l)=\frac{1}{2^\nu}\sum_{\alpha=0}^{i-\nu}a_\alpha
S_{i-\nu-\alpha}(\mbox{\boldmath $x$}+\nu\mbox{\boldmath $s$}),
\]
\[
\Psi_{j\nu}^{(n)}(k,l)=\frac{1}{2^\nu}\sum_{\beta=0}^{j-\nu}\bar a_\beta
S_{j-\nu-\beta}(\mbox{\boldmath $y$}+\nu\mbox{\boldmath $s$}),
\]
$a_\alpha$ are complex constants, $S_\nu(\mbox{\boldmath $x$})$ are
elementary Schur polynomials defined by
\[
\sum_{\nu=0}^{\infty}S_\nu(\mbox{\boldmath $x$})\lambda^\nu
=\exp\left(\sum_{\nu=1}^{\infty}x_\nu\lambda^\nu\right),
\]
$\mbox{\boldmath $x$}=(x_1,x_2,\cdots)$, $\mbox{\boldmath
$y$}=(y_1,y_2,\cdots)$ and $\mbox{\boldmath $s$}=(s_1,s_2,\cdots)$
are defined by
$$
\begin{array}{l}    \hspace{-2.3cm}
x_\nu=(n+k)r_\nu(\rho)+lr_\nu(1/\rho)-r_\nu(1)
+\rho x/\nu!+(1-\rho^2)(\nu+1)r_{\nu+1}(\rho)y-k\delta_{\nu 1},
\\  \hspace{-2.3cm}
y_\nu=(n+l)r_\nu(\rho)+kr_\nu(1/\rho)-r_\nu(1)
+\rho y/\nu!+(1-\rho^2)(\nu+1)r_{\nu+1}(\rho)x-l\delta_{\nu 1},
\end{array}
$$
\begin{equation} \label{rs}
\sum_{\nu=1}^{\infty}r_\nu(\rho)\lambda^\nu
=\ln\frac{1+\rho e^\lambda}{1+\rho},
\qquad
\sum_{\nu=1}^{\infty}s_\nu\lambda^\nu
=\ln\left(\frac{2}{\lambda}\tanh\frac{\lambda}{2}\right),
\end{equation}
$\delta_{\nu 1}$ denotes the Kronecker delta
and $x=ite^{-i\theta}/(1-\rho^2)$, $y=-ite^{i\theta}/(1-\rho^2)$.
The determinant $\sigma_n(k,l)$ can also be expressed as
\begin{eqnarray} \label{sig}
&& \sigma_n(k,l)=\sum_{\nu_1=0}^1\sum_{\nu_2=\nu_1+1}^3
\cdots\sum_{\nu_N=\nu_{N-1}+1}^{2N-1}
\left(\frac{1-\rho}{1+\rho}\right)^{\nu_1+\nu_2+\cdots+\nu_N}  \times \nonumber \\ &&  \hspace{1.8cm}
\det_{1\le i,j\le N}\left(\Phi_{2i-1,\nu_j}^{(n)}(k,l)\right)
\det_{1\le i,j\le N}\left(\Psi_{2i-1,\nu_j}^{(n)}(k,l)\right),
\end{eqnarray}
where we have defined
$\Phi_{i\nu}^{(n)}(k,l)=\Psi_{i\nu}^{(n)}(k,l)=0$ for $i<\nu$.

Regarding boundary conditions of these rogue waves at large times,
we have the following theorem.

\noindent{\textbf{Theorem 3}} \hspace{0.2cm} As $t\to \pm \infty$,
solutions $u_n(t)$ in Theorems 1 and 2 approach a constant
background,
\begin{equation}  \label{e:bc}
u_n(t) \to (-1)^N \frac{\rho}{\sqrt{1-\rho^2}}\hspace{0.15cm} e^{i(\theta n-\omega t)}
\end{equation}
uniformly for all $n$ as long as $\cos\theta\ne 0$.

This theorem confirms that solutions $u_n(t)$ in Theorems 1 and 2
are indeed rogue waves, i.e., they rise from a constant background
and then retreat back to this same background.

Regarding regularity (boundedness) of these rogue waves, we have the
following theorem.

\noindent{\textbf{Theorem 4}} \hspace{0.2cm} General rogue-wave
solutions to the focusing AL equation (\ref{e:ALf}) (with
$|\rho|<1$) in Theorems 1 and 2 are non-singular for all times.

Proofs of these theorems will be presented in section 4.

{\bf Remark 1}\hspace{0.2cm} In these rogue-wave solutions, $\rho$
controls the background amplitude, and $\theta$ is the phase
gradient across the lattice. Obviously, the value of $\theta$ can be
restricted to $-\pi < \theta\le \pi$. Since the AL equations are
invariant with respect to a time shift, we can normalize the
imaginary part of $a_1$ to be zero through a time shift. Then
non-reducible free parameters in these $N$-th order rogue waves are
$\rho, \theta, \mbox{Re}(a_1)$ and $a_3, a_5, \dots a_{2N-1}$,
totaling $2N+1$ real parameters. The parameter $\mbox{Re}(a_1)$ is
equivalent to a shift $n \to n-n_0$ in the solution, with $n_0$
being a real parameter. With this $n$-shift, we can set $a_1=0$. In
this case, $n_0$ becomes a free parameter in the solution instead of
$\mbox{Re}(a_1)$. Without loss of generality, one may restrict
$-1/2<n_0\le 1/2$ through a shift of the lattice index $n$.

{\bf Remark 2} \hspace{0.2cm}  The number of irreducible free
parameters in these rogue waves of the AL equations is three more
than the corresponding number $2N-2$ in the nonlinear Schr\"odinger
(NLS) equation \cite{OY}. The reason is that the NLS equation has
three additional invariances which are lacking in the AL equations:
the spatial-translation invariance, the Galilean-transformation
invariance, and the scaling invariance. These three invariances
reduce the number of free parameters in rogue waves of the NLS
equation by three, thus it is three less than that in the AL
equations. More will be said on this issue in the next section.

{\bf Remark 3}\hspace{0.2cm} It was pointed out in
\cite{Matveev_Kyoto} that the coefficients $s_\nu$ in Eq. (\ref{rs})
are related to Bernoulli numbers $B_\nu$ as
$$
s_\nu=-\frac{2^\nu-2}{\nu!\nu}B_\nu,\quad(\nu\ge2),\qquad s_1=0,
$$
where the Bernoulli numbers $B_\nu$ are defined by
$$
\sum_{\nu=0}^\infty\frac{B_\nu}{\nu!}\lambda^\nu=\frac{\lambda}{e^\lambda-1}.
$$

\section{Dynamics of rogue waves}

In this section, we examine dynamics of rogue waves in AL equations.

\subsection{Fundamental rogue waves}

Fundamental rogue waves are obtained by setting $N=1$ in Eq.
(\ref{f:general}) or (\ref{f:theorem2}). After simple algebra, these
rogue waves are
$$
u_n(t)=-\frac{\rho}{\sqrt{1-\rho^2}} e^{i(\theta n-\omega t)} \left[ 1+\frac{
2i\rho^2\omega t+(1+\rho)(a_1-\bar{a}_1)-1}{\rho^2(1+\rho)^2|R|^2+\frac{1}{4}(1-\rho^2)}\right],
$$
where
\[
R=\frac{1}{1+\rho}\left[n+i\left(\frac{e^{-i\theta}}{1+\rho}-\frac{e^{i\theta}}{1-\rho}\right)t\right]+\frac{\bar{a}_1-1/2}{\rho}.
\]
After shifts of $t$, $n$, and utilizing phase and time-shift
invariances of the AL equations, the above fundamental rogue waves
can be rewritten as
\begin{equation}  \label{f:fundamental}
\hspace{-1.5cm}
u_n(t)=\frac{\rho}{\sqrt{1-\rho^2}} e^{i(\theta n-\omega t)} \left[ 1+\frac{
2i\rho^2\omega t-1}{\rho^2\left(n+\omega t \tan\theta-n_0\right)^2+\rho^4\omega^2t^2+\frac{1}{4}(1-\rho^2)}\right],
\end{equation}
where $\rho, \theta$ and $n_0$ are free real parameters. In view of
Remark 1, we restrict
$$-\pi < \theta\le \pi, \quad -1/2<n_0\le 1/2$$
in this subsection. From the explicit expression
(\ref{f:fundamental}), we see that $|u_n|$ depends on $n$ only
through the combination of $\rho n$, thus $\rho$ controls the
spatial width of this rogue wave (smaller $\rho$ yields broader
waves). Of course, $\rho$ also controls the background amplitude of
this rogue wave. This background amplitude is
\begin{equation} \label{e:r}
r=\frac{|\rho|}{\sqrt{|1-\rho^2|}},
\end{equation}
as is easily seen from Eq. (\ref{f:fundamental}).

Now we compare this fundamental rogue wave (\ref{f:fundamental})
with that reported in \cite{Ankiewicz2010}. There are two main
differences between them. One is that our solution contains one more
free parameter $\theta$ (the phase gradient), whose role will be
explained in the later text. The other difference is that our
solution yields rogue waves for both focusing and defocusing AL
equations, while that in \cite{Ankiewicz2010} only yields rogue
waves for the focusing AL equation.

It is noted that the solution (\ref{f:fundamental}) approaches a
constant background when $t\to \pm \infty$ as long as $\omega\ne 0$,
i.e., $\theta\ne \pm \pi/2$. If $\theta= \pm \pi/2$, then this
solution becomes
$$
u_n(t)=\frac{\rho}{\sqrt{1-\rho^2}} e^{\pm i n\pi/2} \left[ 1-\frac{1}{\rho^2\left(n\pm \frac{2}{1-\rho^2}t-n_0\right)^2+\frac{1}{4}(1-\rho^2)}\right],
$$
which is a soliton moving on a constant background rather than a
rogue wave. For consideration of rogue waves, we will require
$\theta\ne \pm \pi/2$ in the rest of this article.

Dynamics of this rogue wave (\ref{f:fundamental}) differs
significantly for the focusing and defocusing AL equations
(corresponding to $|\rho|<1$ and $|\rho|>1$ respectively). Thus we
will discuss these two cases separately below.

\subsubsection{Focusing case}

In this case, $|\rho|<1$, and Eq. (\ref{f:fundamental}) is the
fundamental rogue wave of the focusing AL equation (\ref{e:ALf}).
Since $|\rho|<1$, the background amplitude (\ref{e:r}) of this rogue
wave can be arbitrary, i.e., $0<r<\infty$. In addition, the
denominator in Eq. (\ref{f:fundamental}) is never zero, thus this
wave is bounded for all time and lattice sites. It is also seen from
Eq. (\ref{f:fundamental}) that $\theta$ can be viewed as a velocity
parameter of this rogue wave, with the velocity being
$-\omega\tan\theta$, i.e.,  $2\sin\theta/(\rho^2-1)$. Thus rogue
waves with $\theta=0, \pi$ can be called stationary, and those with
other $\theta$ values called moving.

Let us first consider stationary rogue waves with $\theta=0$ (the
$\theta=\pi$ case is very similar). At this $\theta$ value,
$\omega=\omega_0\equiv 2/(1-\rho^2)$, and the fundamental rogue wave
(\ref{f:fundamental}) reduces to
\begin{equation}  \label{f:fundamentalzero}
u_n(t)=\frac{\rho}{\sqrt{1-\rho^2}} \hspace{0.1cm}  e^{-i\omega_0 t} \left[ 1+\frac{
2i\rho^2\omega_0 t-1}{\rho^2\left(n-n_0\right)^2+\rho^4\omega_0^2t^2+\frac{1}{4}(1-\rho^2)}\right].
\end{equation}
This rogue wave is equivalent to that reported in
\cite{Ankiewicz2010}. The peak amplitude of this rogue wave is
reached at $t=0$ and the lattice site $n$ which is closest to the
shift parameter $n_0$. The highest peak amplitude occurs when
$n_0=0$. In this case, the highest peak amplitude is
\begin{equation}
u_{\mbox{\scriptsize max}}=\frac{|\rho|}{\sqrt{1-\rho^2}} \hspace{0.1cm} \frac{3+\rho^2}{1-\rho^2}.
\end{equation}
In terms of the background amplitude $r$ defined in Eq. (\ref{e:r}), this highest peak amplitude is
\begin{equation}
u_{\mbox{\scriptsize max}}=r(3+4r^2).
\end{equation}
This peak amplitude is at least three times the background amplitude
$r$, and can be much higher when the background is high. This
amplitude is reached at a single lattice site $n=0$, and can be
called on-site rogue waves.

The lowest peak amplitude of this rogue wave occurs when $n_0=1/2$.
In this case, the peak amplitude is $3r$, which is exactly three
times the background. This peak amplitude is reached at two adjacent
lattice sites $n=0$ and $n=1$ simultaneously and can be called
inter-site rogue waves.

These stationary fundamental rogue waves (\ref{f:fundamentalzero})
are illustrated in Fig. 1. The upper row shows two on-site rogue
waves (with $n_0=0$), and the lower row shows two inter-site rogue
waves (with $n_0=1/2$). On the left column, $\rho=0.2$, which is
small. On the right column, $\rho=0.8$. We can see from this figure
that on-site rogue waves can run much higher than inter-site ones,
especially when $\rho$ is not small (see right column). When $\rho$
is small, rogue waves are broad (see left column). In this case, the
difference between on-site and inter-site waves is less pronounced.

\begin{figure}[h!]
\hspace{2.5cm} \includegraphics[width=0.8\textwidth]{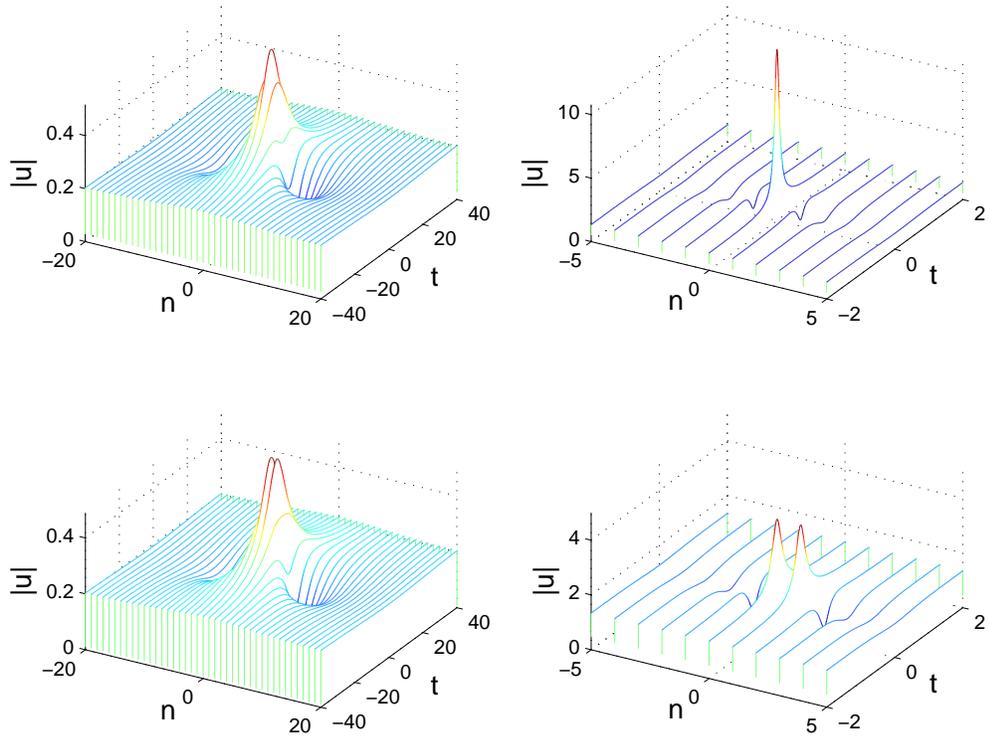}
\caption{Stationary fundamental rogue waves (\ref{f:fundamentalzero}) in the focusing Ablowitz-Ladik equation. Top row: on-site waves ($n_0=0$);
bottom row: inter-site waves ($n_0=1/2$); left column: broad waves ($\rho=0.2$); right column: narrow waves ($\rho=0.8$). }
\end{figure}

Now we consider moving rogue waves (\ref{f:fundamental}) with
$\theta \ne 0, \pi$. These rogue waves have not been reported before
\cite{Ankiewicz2010}. Two such solutions, with $\rho=0.8, n_0=0$ and
$\theta=-1.2, -0.2$ are displayed in Fig. 2. In the left figure, we
see a rogue wave rising from the constant background, traversing
across the lattice, and then disappearing into the background again.
In the right figure, the traversing motion of the rogue wave is less
visible, because this rogue wave rises to its peak amplitude and
retreats back to the constant background more quickly.

\begin{figure}[h!]
\hspace{2.5cm} \includegraphics[width=0.8\textwidth]{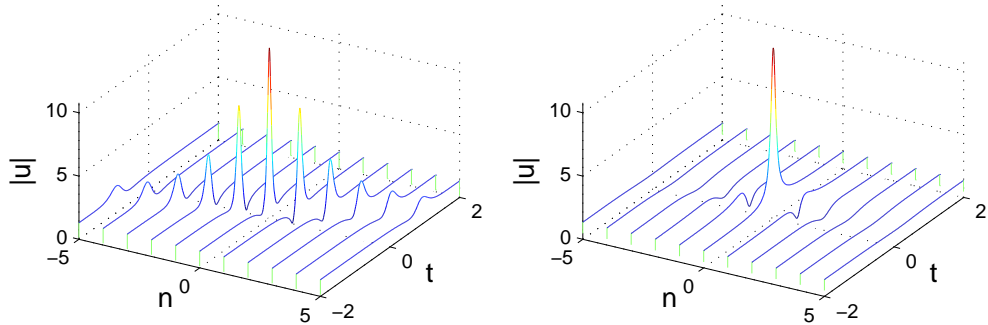}
\caption{Moving fundamental rogue waves (\ref{f:fundamental}) in the focusing Ablowitz-Ladik equation with $\rho=0.8$ and $n_0=0$.
Left: $\theta=-1.2$; right: $\theta=-0.2$. }
\end{figure}

When $\rho \to 0$, fundamental rogue waves (\ref{f:fundamental})
become very broad. In this case, the AL equation (\ref{e:ALf})
reduces to the NLS equation, and rogue waves (\ref{f:fundamental})
approach the fundamental rogue waves of the NLS equation. In this
limit, the parameter $\theta$ is the counterpart of the moving
velocity of NLS rogue waves. The NLS equation admits Galilean
invariance, thus any moving rogue wave can be derived from a
stationary one through Galilean transformation. However, the AL
equation is not Galilean invariant. Because of that, $\theta$ is a
non-reducible parameter in rogue waves of the AL equation.

\subsubsection{Defocusing case}

Now we consider the defocusing case, where $|\rho|>1$, and solution
(\ref{f:fundamental}) satisfies the defocusing AL equation
(\ref{e:ALd}). In this case, solution (\ref{f:fundamental}) still
approaches the constant background as $t \to \pm\infty$, and rises
to higher amplitude in the intermediate times, thus is also a rogue
wave. The existence of rogue waves in the defocusing AL equation is
surprising. Notice that the background amplitude $r$ of these rogue
waves is always larger than 1 since $|\rho|>1$ [see Eq.
(\ref{e:r})]. Indeed, we can show that in the defocusing AL
equation, rogue waves with background amplitudes less than 1 cannot
exist since such backgrounds are modulationally stable (see next
subsection).

Rogue waves in the defocusing AL equation exhibit new features that
have no counterparts in the focusing AL equation. Since $|\rho|>1$,
the denominator in Eq. (\ref{f:fundamental}) may become zero, thus
this rogue wave may blow up to infinity in finite time. To
illustrate, let us take $\theta=0$. Then from the explicit
expression (\ref{f:fundamentalzero}), we see that this rogue wave
will explode to infinity if
\begin{equation}  \label{e:condblowup}
|n_0|<\frac{1}{2r},
\end{equation}
where $r$ is the background amplitude in Eq. (\ref{e:r}). When
$n_0=1/2$, this rogue wave will be a regular rogue wave and never
blow up since $r>1$. An example is shown in Fig. 3 (left) with
$\rho=2$. This is an inter-site rogue wave, resembling that in Fig.
1 (lower right panel) of the focusing AL equation. However, if
$n_0=0$, then the rogue wave (\ref{f:fundamentalzero}) will always
blow up. An example is shown in Fig. 3 (right) with $\rho=2$. We see
that this rogue wave blows up to infinity at the lattice site $n=0$
and times $t=\pm 3\sqrt{3}/16$. At other $n_0$ values of
$0<|n_0|<1/2$, this rogue wave will blow up for background values
determined by the condition (\ref{e:condblowup}). The $n_0$-range
for wave blowup shrinks as the background amplitude increases.

\begin{figure}[h!]
\hspace{2.5cm} \includegraphics[width=0.8\textwidth]{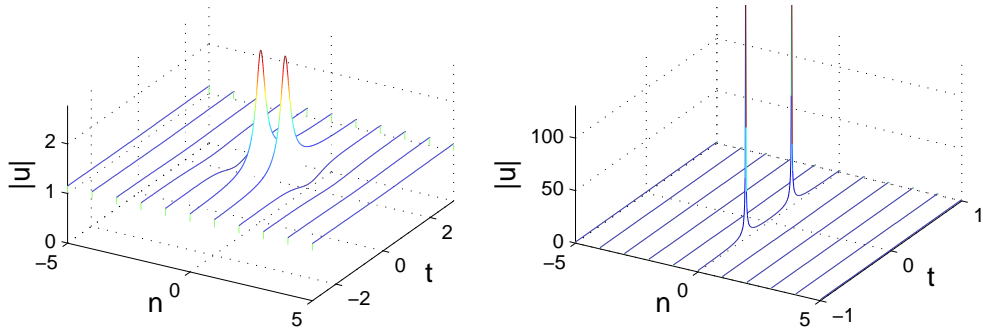}
\caption{Fundamental rogue waves (\ref{f:fundamental}) in the defocusing Ablowitz-Ladik equation with $\theta=0$ and $\rho=2$. Left: $n_0=1/2$; right: $n_0=0$. }
\end{figure}

One may recall that exploding rogue waves have also been reported
for the Davey-Stewartson (DS) II equation \cite{OY_DSII}. However,
blowup in the DSII equation appears only for second- and
higher-order rogue waves, but the blowup here occurs even for
fundamental rogue waves (with $N=1$).

\subsubsection{Connection with modulation instability }

Why do rogue waves with background amplitudes higher than 1 exist in
the defocusing AL equation? The reason is that such backgrounds are
modulationally unstable. This modulation instability is analyzed
below.

The defocusing AL equation (\ref{e:ALd}) admits a
constant-background solution
\begin{equation}
u_n(t)=r \hspace{0.03cm} e^{-2i(1-r^2)t},
\end{equation}
where $r$ is the background amplitude. To study the modulation
instability of this constant-background solution, we perturb this
solution by normal modes
\begin{equation} \label{e:unperturb}
u_n(t)=e^{-2i(1-r^2)t} \left(r + f e^{\lambda t+i\beta n} +\bar{g} e^{\bar{\lambda} t-i\beta n}\right),
\end{equation}
where $\lambda$ and $\beta$ are the growth rate and wavenumber of
the perturbation, and $f, g \ll 1$.
Substituting this perturbed solution in Eq. (\ref{e:ALd}) and
neglecting terms of higher order in $f$ and $g$, we obtain the
following equation for the growth rate $\lambda$,
\begin{equation} \label{f:lambda2}
\lambda^2=4(r^2-1)(1-\cos\beta)\left[(r^2+1)+(r^2-1)\cos\beta\right].
\end{equation}
This formula shows that when the background amplitude $r>1$,
$\lambda^2$ is positive for all wavenumbers $\beta$ with
$\cos\beta\ne 1$, thus this constant background is modulationally
unstable. As a consequence, rogue waves with background amplitudes
higher than 1 can exist in the defocusing AL equation (\ref{e:ALd}).

For lower background amplitudes $0<r<1$, however, the formula
(\ref{f:lambda2}) shows that $\lambda^2$ is never positive for any
wavenumber $\beta$, thus backgrounds lower than 1 are modulationally
stable in the defocusing AL equation. Consequently rogue waves with
such lower backgrounds cannot exist.

This modulation stability analysis can also be performed for the
focusing AL equation (\ref{e:ALf}). In this case, the
constant-background solution is
$$
u_n(t)=r \hspace{0.03cm} e^{-2i(1+r^2)t}.
$$
Perturbing this solution by normal modes similar to
(\ref{e:unperturb}) and following similar procedures, we can obtain
the following equation for the growth rate $\lambda$,
$$
\lambda^2=4(r^2+1)(1-\cos\beta)\left[(r^2-1)+(r^2+1)\cos\beta\right].
$$
This formula shows that, for any background amplitude $r$,
$\lambda^2$ is positive for wavenumbers $\beta$ with $\cos\beta >
(1-r^2)/(1+r^2)$; thus all constant backgrounds in the focusing AL
equation are modulationally unstable. This explains why rogue waves
with arbitrary constant backgrounds exist in the focusing AL
equation (\ref{e:ALf}).

\subsection{Second-order rogue waves}

Now we consider second-order rogue waves in the AL equations. These
second-order rogue waves can be obtained from formula
(\ref{f:general}) or (\ref{f:theorem2}) by taking
$$ N=2, \;  a_1=a_2=0, $$
and shifting $n$ to $n-n_0$, with $n_0$, $\theta$, $\rho$ and $a_3$
being free parameters. For simplicity, we take $\theta=0$ in our
discussions below.

\subsubsection{ Focusing case }

In this case, $|\rho|<1$. For $\rho=1/2$, four second-order rogue
waves are displayed in Fig. 4 (the $n_0$ and $a_3$ parameters are
specified in the captions). We see that these second-order rogue
waves are all bounded (no blowup). In addition, they can exhibit
either a single dominant hump (see panel (a)), or three humps (see
panels (b-d)), depending on parameters. These behaviors are
analogous to second-order rogue waves of the NLS equation
\cite{Rogue_higher_order,Rogue_higher_order2,Rogue_Gaillard,Rogue_triplet,Rogue_circular,Liu_qingping,
OY,Matveev_Kyoto}. However, differences between AL and NLS rogue
waves are also apparent. The main difference is that the three humps
of the AL rogue waves generally have different heights, while those
of the NLS rogue waves generally have the same height. The reason
for this difference is that, in the AL rogue waves, some of these
three humps are on-site and the others inter-site. On-site humps
have higher heights than inter-site ones (see the previous
subsection).

\begin{figure}[h!]
\hspace{2.5cm} \includegraphics[width=0.8\textwidth]{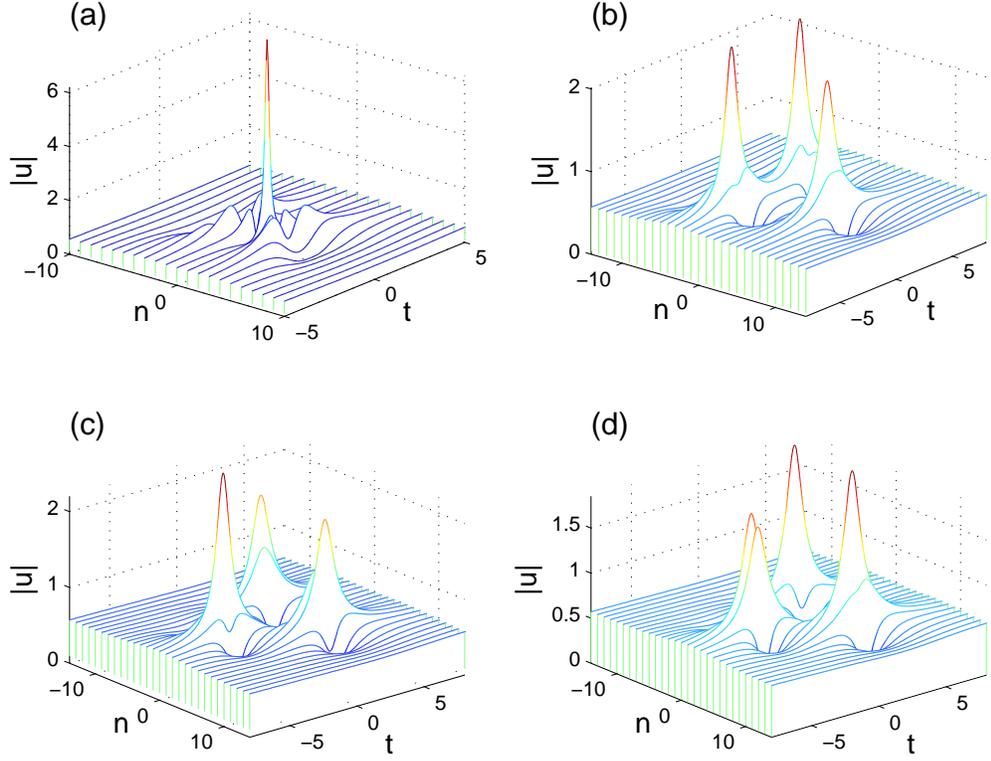}
\caption{Second-order rogue waves in the focusing Ablowitz-Ladik equation for $\rho=1/2$ and $\theta=0$. (a) $n_0=-3/2, a_3=-1/54$; (b) $n_0=-3/2, a_3=5/3$;
(c) $n_0=-3/2, a_3=5i/3$; (d) $n_0=0, a_3=5i/3$.  }
\end{figure}

Second-order rogue waves in the focusing AL equation have been
reported before \cite{Ankiewicz2010}. Those rogue waves contain only
two free real parameters (the counterparts of $\rho$ and $n_0$ in
this article), thus they are a special class of second-order rogue
waves. Due to the lack of the complex free parameter $a_3$, those
second-order rogue waves in \cite{Ankiewicz2010} cannot exhibit
three-hump structures such as Fig. 4(b-d).

For each given $|\rho|<1$, we have also explored the rogue wave with
the highest peak amplitude among all second-order rogue waves with
free $n_0$ and $a_3$ values. We find that the highest possible peak
amplitude is
\begin{equation}  \label{e:high2}
|u|_{\mbox{\scriptsize max}}=r(5+20r^2+16r^4),
\end{equation}
where $r=|\rho|/\sqrt{|1-\rho^2|}$ is the background amplitude [see
Eq. (\ref{e:r})]. Interestingly, this highest-amplitude formula is
identical to that reported in \cite{Ankiewicz2010} even though the
second-order rogue waves obtained in that work were special.

In this rogue wave with the highest peak amplitude (\ref{e:high2}),
the corresponding $n_0$ and $a_3$ values are
\begin{equation}
n_{0, \hspace{0.05cm} \mbox{\scriptsize max}}=-\frac{1+\rho}{2\rho}, \quad a_{3, \hspace{0.05cm} \mbox{\scriptsize max}}=\frac{1}{12} \frac{\rho-1}{(\rho+1)^2},
\end{equation}
and this peak amplitude occurs at
\begin{equation}
n_{\mbox{\scriptsize max}}=0, \quad t_{\mbox{\scriptsize max}}=0.
\end{equation}
For $\rho=1/2$, $n_{0, \hspace{0.04cm} \mbox{\scriptsize
max}}=-3/2$,  and $a_{3, \hspace{0.04cm} \mbox{\scriptsize
max}}=-1/54$. The corresponding rogue wave is as displayed in Fig.
4(a). This rogue wave reaches peak amplitude $121/(9\sqrt{3})$,
which is about 13 times higher than the background amplitude
$1/\sqrt{3}$.

\subsubsection{ Defocusing case }

Next we consider second-order rogue waves in the defocusing AL
equation, where $|\rho|>1$. For $\rho=1.2$, four of these rogue
waves are displayed in Fig. 5. We see that these second-order rogue
waves can be bounded for certain parameter values (see panels
(a,b)). However, for many other parameter values, they blow up in
finite time (see panels (c,d)). This existence of both bounded and
exploding second-order rogue waves is similar to that in fundamental
rogue waves of the defocusing AL equation (see Fig. 3).

\begin{figure}[h!]
\hspace{2.5cm} \includegraphics[width=0.8\textwidth]{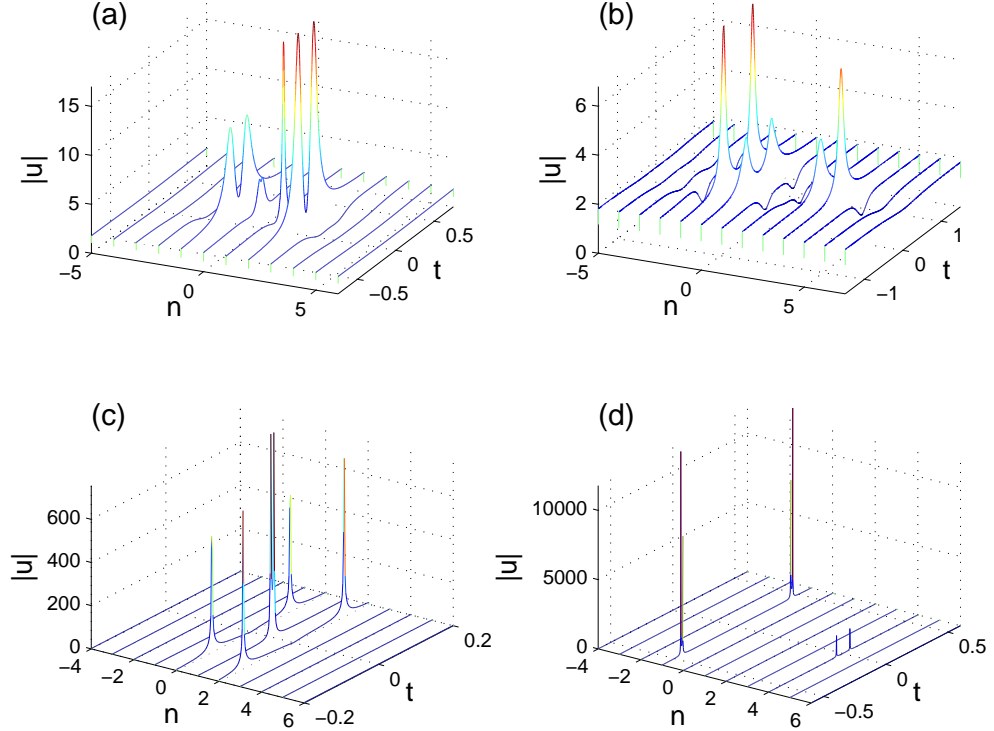}
\caption{Second-order rogue waves in the defocusing Ablowitz-Ladik equation for $\rho=1.2$ and $\theta=0$.
Top row: bounded rogue waves; bottom row: exploding rogue waves. The solution parameters are
(a) $a_3=0, n_0=-0.38$; (b) $a_3=5/6, n_0=0$; (c) $a_3=0, n_0=0$; (d)
$a_3=4/3, n_0=0$.   }
\end{figure}

\subsection{Higher-order rogue waves}

Dynamics of third and higher order rogue waves in the AL equations
can be studied in a similar way by using the general formula
(\ref{f:general}) or (\ref{f:theorem2}). For instance, we consider
third-order rogue waves in the focusing AL equation by taking
$$ N=3, \;  \theta=0, \;  \rho=1/2, \; a_1=a_2=a_4=n_0=0, $$
with $a_3$ and $a_5$ as free parameters. For four choices of ($a_3,
a_5$) values, the corresponding rogue waves are displayed in Fig.~6.
It is seen that this rogue wave can exhibit a single high peak, or
six lower peaks, depending on the ($a_3, a_5$) values. Notice that
the six peaks in Fig.~6 form triangular or circular patterns,
analogous to the NLS equation \cite{Rogue_circular,Liu_qingping,
OY}. However, the six peaks here have uneven amplitudes, unlike the
NLS equation where the six peaks have almost identical amplitudes.

\begin{figure}[h!]
\hspace{2.5cm} \includegraphics[width=0.8\textwidth]{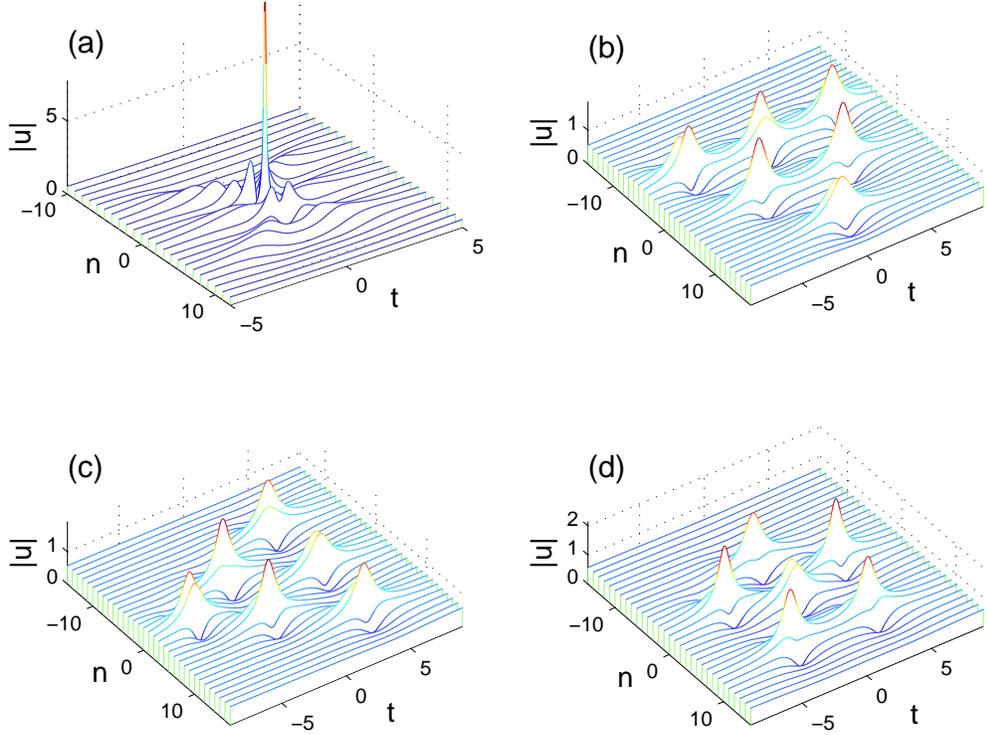}
\caption{Third-order rogue waves in the focusing Ablowitz-Ladik equation for $\rho=1/2$, $\theta=0$ and $n_0=0$. (a) $a_3=a_5=0$; (b) $a_3=2, a_5=0$;
(c) $a_3=2i, a_5=0$; (d) $a_3=0, a_5=2$. }
\end{figure}

\section{Derivation of rogue-wave solutions}
In this section, we derive general rogue-wave solutions and prove
their boundary and regularity properties in Theorems 1-4.

We first establish a few lemmas. In Lemma 1 we introduce the
so-called Grammian solutions to certain bilinear
differential-difference equations, which are relevant to our study.
By assuming that the matrix elements obey appropriate dispersion
relations, we can show that the determinant (which we call $\tau$
function) satisfies these bilinear equations. If we choose suitable
matrix elements, this $\tau$ function gives polynomial solutions,
which is explained in Lemma 2. The next crucial step is to apply
reduction to these polynomial solutions. This reduction is achieved
in Lemma 3. Then by constraining parameters in the matrix elements,
the $\tau$ function satisfies certain reality and conjugacy
conditions, hence its bilinear equations reduce to the AL equations
through a variable transformation. Rogue waves in the AL equations
then are expressed through this $\tau$ function.

\noindent{\bf Lemma 1}\hspace{0.2cm} Let $m_{ij}^{(n)}$,
$\varphi_i^{(n)}$ and $\psi_j^{(n)}$ be functions of continuous
independent variables $x$, $y$ and discrete ones $k$, $l$,
satisfying the following differential and difference (dispersion)
relations,
\begin{equation} \label{ddrel}
\begin{array}{l}
\partial_xm_{ij}^{(n)}(k,l)=\varphi_i^{(n)}(k,l)\psi_j^{(n-1)}(k,l),
\\[5pt]
\partial_ym_{ij}^{(n)}(k,l)=\varphi_i^{(n-1)}(k,l)\psi_j^{(n)}(k,l),
\\[5pt]
m_{ij}^{(n+1)}(k,l)=(1-\rho^2)m_{ij}^{(n)}(k,l)
+\varphi_i^{(n)}(k,l)\psi_j^{(n)}(k,l),
\\[5pt]
m_{ij}^{(n)}(k+1,l)=(1-\rho^2)m_{ij}^{(n)}(k,l)
-\varphi_i^{(n-1)}(k+1,l)\psi_j^{(n)}(k,l),
\\[5pt]
m_{ij}^{(n)}(k,l+1)=(1-\rho^2)m_{ij}^{(n)}(k,l)
-\varphi_i^{(n)}(k,l)\psi_j^{(n-1)}(k,l+1),
\\[5pt]
\partial_x\varphi_i^{(n)}(k,l)=\varphi_i^{(n+1)}(k,l),
\\[5pt]
\partial_y\varphi_i^{(n)}(k,l)=-(1-\rho^2)\varphi_i^{(n-1)}(k,l),
\\[5pt]
\varphi_i^{(n)}(k-1,l)=\varphi_i^{(n)}(k,l)-\varphi_i^{(n-1)}(k,l),
\\[5pt]
\varphi_i^{(n)}(k,l+1)=(1-\rho^2)\varphi_i^{(n)}(k,l)-\varphi_i^{(n+1)}(k,l),
\\[5pt]
(\partial_x+1)\psi_j^{(n)}(k,l)=-\psi_j^{(n-1)}(k+1,l),
\\[5pt]
(\partial_y-1)\psi_j^{(n)}(k,l)=\psi_j^{(n+1)}(k,l-1),
\\[5pt]
\psi_j^{(n)}(k+1,l)=(1-\rho^2)\psi_j^{(n)}(k,l)-\psi_j^{(n+1)}(k,l),
\\[5pt]
\psi_j^{(n)}(k,l-1)=\psi_j^{(n)}(k,l)-\psi_j^{(n-1)}(k,l).
\end{array}
\end{equation}
Then the determinant
$$
\tau_n(k,l)=\det_{1\le i,j\le N}\left(m_{ij}^{(n)}(k,l)\right)
$$
satisfies the bilinear equations
\begin{equation} \label{bilin}
\begin{array}{l}
(D_x+1)\tau_n(k-1,l)\cdot\tau_n(k,l)=\tau_{n+1}(k-1,l)\tau_{n-1}(k,l),
\\
(D_y-1)\tau_n(k,l+1)\cdot\tau_n(k,l)=-\tau_{n-1}(k,l+1)\tau_{n+1}(k,l),
\\
\tau_{n+1}(k-1,l)\tau_{n-1}(k,l+1)-(1-\rho^2)\tau_n(k-1,l)\tau_n(k,l+1)  \\
\hspace{5.2cm} =\rho^2\tau_n(k-1,l+1)\tau_n(k,l).
\end{array}
\end{equation}
Here $D$ is Hirota's bilinear differential operator defined by
\begin{eqnarray*}
\hspace{-1cm} P(D_x,D_y)F(x,y)\cdot G(x,y)=P(\partial_x-\partial_{x'},\partial_y-\partial_{y'})
F(x,y)G(x',y')|_{x'=x,y'=y},
\end{eqnarray*}
and $P$ is a polynomial of $D_x$ and $D_y$.

\noindent{\bf Proof}\hspace{0.2cm} By using the dispersion relations
(\ref{ddrel}), we can show that the derivatives and shifts of the
$\tau$ function are expressed by the bordered determinants as
follows,
\begin{eqnarray*}
&&\hspace{-1.5cm}  \partial_x\tau_n(k,l)=\left|\matrix{
m_{ij}^{(n)}(k,l) &\varphi_i^{(n)}(k,l) \cr
-\psi_j^{(n-1)}(k,l) &0}\right|,
\\
&&\hspace{-1.5cm} \partial_y\tau_n(k,l)=\left|\matrix{
m_{ij}^{(n)}(k,l) &\varphi_i^{(n-1)}(k,l) \cr
-\psi_j^{(n)}(k,l) &0}\right|,
\\
&& \hspace{-1.5cm} \tau_{n+1}(k,l)=(1-\rho^2)^{N-1}\left|\matrix{
m_{ij}^{(n)}(k,l) &\varphi_i^{(n)}(k,l) \cr
-\psi_j^{(n)}(k,l) &1-\rho^2}\right|,
\\
&&\hspace{-1.5cm} \tau_{n-1}(k,l)=\frac{1}{(1-\rho^2)^N}\left|\matrix{
m_{ij}^{(n)}(k,l) &\varphi_i^{(n-1)}(k,l) \cr
\psi_j^{(n-1)}(k,l) &1}\right|,
\\
&&\hspace{-1.5cm} \tau_n(k-1,l)=\frac{1}{(1-\rho^2)^N}\left|\matrix{
m_{ij}^{(n)}(k,l) &\varphi_i^{(n-1)}(k,l) \cr
-\psi_j^{(n)}(k-1,l) &1}\right|,
\\
&&\hspace{-1.5cm} \tau_n(k,l+1)=(1-\rho^2)^{N-1}\left|\matrix{
m_{ij}^{(n)}(k,l) &\varphi_i^{(n)}(k,l) \cr
\psi_j^{(n-1)}(k,l+1) &1-\rho^2}\right|,
\\
&&\hspace{-1.5cm}  (\partial_x+1)\tau_n(k-1,l)=\frac{1}{(1-\rho^2)^N}\left|\matrix{
m_{ij}^{(n)}(k,l) &\varphi_i^{(n-1)}(k,l) &\varphi_i^{(n)}(k,l) \cr
-\psi_j^{(n)}(k-1,l) &1 &1 \cr
-\psi_j^{(n-1)}(k,l) &-1 &0}\right|,
\\
&&\hspace{-1.5cm}  (\partial_y-1)\tau_n(k,l+1)=(1-\rho^2)^{N-1}\left|\matrix{
m_{ij}^{(n)}(k,l) &\varphi_i^{(n)}(k,l) &\varphi_i^{(n-1)}(k,l) \cr
\psi_j^{(n-1)}(k,l+1) &1-\rho^2 &1 \cr
-\psi_j^{(n)}(k,l) &1-\rho^2 &0}\right|,
\\
&&\hspace{-1.5cm}  \tau_{n+1}(k-1,l)=\left|\matrix{
m_{ij}^{(n)}(k,l) &\varphi_i^{(n)}(k,l) \cr
-\psi_j^{(n)}(k-1,l) &1}\right|,
\\
&&\hspace{-1.5cm}  \tau_{n-1}(k,l+1)=\left|\matrix{
m_{ij}^{(n)}(k,l) &\varphi_i^{(n-1)}(k,l) \cr
\psi_j^{(n-1)}(k,l+1) &1}\right|,
\\
&&\hspace{-1.5cm}  \tau_n(k-1,l+1)=\frac{1}{\rho^2}\left|\matrix{
m_{ij}^{(n)}(k,l) &\varphi_i^{(n)}(k,l) &\varphi_i^{(n-1)}(k,l) \cr
-\psi_j^{(n)}(k-1,l) &1 &1 \cr
\psi_j^{(n-1)}(k,l+1) &1-\rho^2 &1}\right|.
\end{eqnarray*}
By using the Jacobi formula of determinants, we obtain the identities,
\begin{eqnarray*}
&& \hspace{-1.5cm} [(\partial_x+1)\tau_n(k-1,l)]\tau_n(k,l)
=\tau_n(k-1,l)\partial_x\tau_n(k,l)+\tau_{n+1}(k-1,l)\tau_{n-1}(k,l),
\\
&& \hspace{-1.5cm} [(\partial_y-1)\tau_n(k,l+1)]\tau_n(k,l)
=\tau_n(k,l+1)\partial_y\tau_n(k,l)-\tau_{n-1}(k,l+1)\tau_{n+1}(k,l),
\\
&& \hspace{-1.5cm} \rho^2\tau_n(k-1,l+1)\tau_n(k,l)  \\ && \hspace{-1.5cm}
=\tau_{n+1}(k-1,l)\tau_{n-1}(k,l+1)-(1-\rho^2)\tau_n(k-1,l)\tau_n(k,l+1),
\end{eqnarray*}
which completes the proof of bilinear equations (\ref{bilin}).
\hfill\fbox{}

The above lemma is quite powerful for constructing various types of
solutions to bilinear equations (\ref{bilin}), since the matrix
elements can be any functions satisfying the dispersion relations
(\ref{ddrel}). For example, a class of polynomial solutions can be
obtained from it by the choice of matrix elements (see next lemma).

\noindent{\bf Lemma 2}\hspace{0.2cm}  We define matrix elements
$m_{ij}^{(n)}$ by
$$
m_{ij}^{(n)}(k,l)=A_iB_jm^{(n)}(k,l),
$$
where
$$
m^{(n)}(k,l)=\frac{1}{pq-1+\rho^2}(pq)^n
\left(\frac{1-\rho^2-q}{1-1/p}\right)^k
\left(\frac{1-\rho^2-p}{1-1/q}\right)^le^{\xi+\eta},
$$
$$
\xi=px-\frac{1-\rho^2}{p}y, \quad \eta=-\frac{1-\rho^2}{q}x+qy,
$$
$A_i$ and $B_j$ are differential operators with respect to $p$ and
$q$ respectively, defined as
\begin{equation}  \label{d:Ai2}
A_i=\sum_{\nu=0}^i\frac{a_{\nu}}{(i-\nu)!}[(p-1)\partial_p]^{i-\nu},
\end{equation}
\begin{equation}  \label{d:Bj2}
B_j=\sum_{\mu=0}^j\frac{b_{\mu}}{(j-\mu)!}[(q-1)\partial_q]^{j-\mu},
\end{equation}
and $a_{\nu}$, $b_{\mu}$ are constants. Then for any sequences of
indices $I_1$, $I_2$, $\cdots$, $I_N$ and $J_1$, $J_2$, $\cdots$,
$J_N$, the determinant,
$$
\tau_n(k,l)=\det_{1\le i,j \le N}\left(m_{I_i,J_j}^{(n)}(k,l)\right)
$$
satisfies the bilinear equations (\ref{bilin}).

\noindent{\bf Proof} \hspace{0.2cm} It is easy to see that the above
$m^{(n)}(k,l)$ and
$$
\varphi^{(n)}(k,l)=p^n(1-1/p)^{-k}(1-\rho^2-p)^le^{\xi},
$$
$$
\psi^{(n)}(k,l)=q^n(1-\rho^2-q)^k(1-1/q)^{-l}e^{\eta}
$$
satisfy the following differential and difference relations,
\begin{eqnarray*}
&&\partial_xm^{(n)}(k,l)=\varphi^{(n)}(k,l)\psi^{(n-1)}(k,l),
\\
&&\partial_ym^{(n)}(k,l)=\varphi^{(n-1)}(k,l)\psi^{(n)}(k,l),
\\
&&m^{(n+1)}(k,l)=(1-\rho^2)m^{(n)}(k,l)
+\varphi^{(n)}(k,l)\psi^{(n)}(k,l),
\\
&&m^{(n)}(k+1,l)=(1-\rho^2)m^{(n)}(k,l)
-\varphi^{(n-1)}(k+1,l)\psi^{(n)}(k,l),
\\
&&m^{(n)}(k,l+1)=(1-\rho^2)m^{(n)}(k,l)
-\varphi^{(n)}(k,l)\psi^{(n-1)}(k,l+1),
\\
&&\partial_x\varphi^{(n)}(k,l)=\varphi^{(n+1)}(k,l),
\\
&&\partial_y\varphi^{(n)}(k,l)=-(1-\rho^2)\varphi^{(n-1)}(k,l),
\\
&&\varphi^{(n)}(k-1,l)=\varphi^{(n)}(k,l)-\varphi^{(n-1)}(k,l),
\\
&&\varphi^{(n)}(k,l+1)
=(1-\rho^2)\varphi^{(n)}(k,l)-\varphi^{(n+1)}(k,l),
\\
&&(\partial_x+1)\psi^{(n)}(k,l)=-\psi^{(n-1)}(k+1,l),
\\
&&(\partial_y-1)\psi^{(n)}(k,l)=\psi^{(n+1)}(k,l-1),
\\
&&\psi^{(n)}(k+1,l)=(1-\rho^2)\psi^{(n)}(k,l)-\psi^{(n+1)}(k,l),
\\
&&\psi^{(n)}(k,l-1)=\psi^{(n)}(k,l)-\psi^{(n-1)}(k,l).
\end{eqnarray*}
Thus
$$
m_{ij}^{(n)}(k,l)=A_iB_jm^{(n)}(k,l), \hspace{0.2cm}
\varphi_i^{(n)}(k,l)=A_i\varphi^{(n)}(k,l), \hspace{0.2cm}
\psi_j^{(n)}(k,l)=B_j\psi^{(n)}(k,l)
$$
satisfy the dispersion relations (\ref{ddrel}). Consequently the
determinant $\tau_n(k,l)$ satisfies the bilinear equations
(\ref{bilin}). This completes the proof. \hfill\fbox{}

We note that the above $\tau_n(k,l)$ is not just a polynomial in
$(x,y,k,l,n)$ but a polynomial times the exponential of a linear
function, because the elements $m_{ij}^{(n)}(k,l)$ have the form of
$(i+j)$-th degree polynomial of $(x,y,k,l,n)$ times
$(pq)^n\left[(1-\rho^2-q)/(1-1/p)\right]^k
\left[(1-\rho^2-p)/(1-1/q)\right]^le^{\xi+\eta}$. The bilinear
equations (\ref{bilin}) are invariant when multiplying an
exponential factor of a linear function in $(x,y,k,l,n)$ to
$\tau_n(k,l)$. Thus through this gauge invariance, the solutions in
Lemma~2 are equivalent to polynomial solutions. In this class of
polynomials, there is a subclass of solutions which satisfy a
certain reduction condition, which is described in the following
lemma.

\noindent{\bf Lemma 3}\hspace{0.2cm} The determinant
\begin{equation}  \label{e:taulemma3}
\tau_n(k,l)=\left.\det_{1\le i,j\le N}\left(m_{2i-1,2j-1}^{(n)}(k,l)\right)
\right|_{p=q=1+\rho},
\end{equation}
where $m_{ij}^{(n)}(k,l)$ is defined in Lemma 2, satisfies the
reduction condition
\begin{equation} \label{red}
\tau_n(k+1,l+1)=(1+\rho)^{4N}\tau_n(k,l).
\end{equation}

\noindent{\bf Proof}\hspace{0.2cm}  We have
\begin{eqnarray*}
&&m^{(n)}(k+1,l+1)
=\frac{1-\rho^2-q}{1-1/p}\ \frac{1-\rho^2-p}{1-1/q}\ m^{(n)}(k,l)
\\
&&\qquad
=\left(p+\rho^2+\frac{\rho^2}{p-1}\right)
\left(q+\rho^2+\frac{\rho^2}{q-1}\right)m^{(n)}(k,l).
\end{eqnarray*}
 From the general Leibniz rule for high-order derivatives of a
product function, we get the operator identity
\begin{eqnarray*}
&& [(p-1)\partial_p]^{\nu}\left(p-1+1+\rho^2+\frac{\rho^2}{p-1}\right)  \\  &&
=\sum_{\kappa=0}^{\nu}{\nu \choose \kappa}
\left(p-1+\delta_{\kappa 0}(1+\rho^2)
+(-1)^{\kappa}\frac{\rho^2}{p-1}\right)[(p-1)\partial_p]^{\nu-\kappa}.
\end{eqnarray*}
Using this
identity, we get
\begin{eqnarray*}
&&\hspace{-1.5cm}  A_i\left(p+\rho^2+\frac{\rho^2}{p-1}\right)  \\&&\hspace{-1.5cm}
=\sum_{\nu=0}^i
\sum_{\kappa=0}^{i-\nu}\frac{a_{\nu}}{\kappa!(i-\nu-\kappa)!}
\left(p-1+\delta_{\kappa 0}(1+\rho^2)
+(-1)^{\kappa}\frac{\rho^2}{p-1}\right)[(p-1)\partial_p]^{i-\nu-\kappa}
\\
&&\hspace{-1.5cm}
=\sum_{\kappa=0}^i\frac{1}{\kappa!}
\left(p-1+\delta_{\kappa 0}(1+\rho^2)
+(-1)^{\kappa}\frac{\rho^2}{p-1}\right)
\sum_{\nu=0}^{i-\kappa}\frac{a_{\nu}}{(i-\nu-\kappa)!}
[(p-1)\partial_p]^{i-\nu-\kappa}
\\
&&\hspace{-1.5cm}
=\sum_{\kappa=0}^i\frac{1}{\kappa!}
\left(p-1+\delta_{\kappa 0}(1+\rho^2)
+(-1)^{\kappa}\frac{\rho^2}{p-1}\right)A_{i-\kappa},
\end{eqnarray*}
and similarly
$$
B_j\left(q+\rho^2+\frac{\rho^2}{q-1}\right)
=\sum_{\lambda=0}^j\frac{1}{\lambda!}
\left(q-1+\delta_{\lambda 0}(1+\rho^2)
+(-1)^{\lambda}\frac{\rho^2}{q-1}\right)B_{j-\lambda}.
$$
Thus the matrix elements satisfy the relation
\begin{eqnarray*}
&& \hspace{-1cm}
m_{ij}^{(n)}(k+1,l+1)=A_iB_j
\left(p+\rho^2+\frac{\rho^2}{p-1}\right)
\left(q+\rho^2+\frac{\rho^2}{q-1}\right)m^{(n)}(k,l)
\\
&& \hspace{-1cm}
=\sum_{\kappa=0}^i\frac{1}{\kappa!}
\left(p-1+\delta_{\kappa 0}(1+\rho^2)
+(-1)^{\kappa}\frac{\rho^2}{p-1}\right)
\\
&& \hspace{-1cm} \times
\sum_{\lambda=0}^j\frac{1}{\lambda!}
\left(q-1+\delta_{\lambda 0}(1+\rho^2)
+(-1)^{\lambda}\frac{\rho^2}{q-1}\right)m_{i-\kappa,j-\lambda}^{(n)}(k,l).
\end{eqnarray*}
Substituting $p=1+\rho$ and $q=1+\rho$, we obtain the contiguity
relation
\begin{eqnarray*}
&&  \hspace{-1cm}
\left. m_{ij}^{(n)}(k+1,l+1)\right|_{p=q=1+\rho}
\\&& \hspace{-1cm}
=\sum_{\scriptstyle\kappa=0 \atop \scriptstyle\kappa:{\rm even}}^i
\frac{2\rho+\delta_{\kappa 0}(1+\rho^2)}{\kappa!}
\sum_{\scriptstyle\lambda=0 \atop \scriptstyle\lambda:{\rm even}}^j
\frac{2\rho+\delta_{\lambda 0}(1+\rho^2)}{\lambda!}
\left.m_{i-\kappa,j-\lambda}^{(n)}(k,l)\right|_{p=q=1+\rho},
\end{eqnarray*}
from which the following matrix relation is derived:
\begin{eqnarray*}
&& \hspace{-2.5cm} \pmatrix{
m_{11}^{(n)}(k+1,l+1) &m_{13}^{(n)}(k+1,l+1)
&\cdots &m_{1,2N-1}^{(n)}(k+1,l+1) \cr
m_{31}^{(n)}(k+1,l+1) &m_{33}^{(n)}(k+1,l+1)
&\cdots &m_{3,2N-1}^{(n)}(k+1,l+1) \cr
\vdots &\vdots &\ddots &\vdots \cr
m_{2N-1,1}^{(n)}(k+1,l+1) &m_{2N-1,3}^{(n)}(k+1,l+1)
&\cdots &m_{2N-1,2N-1}^{(n)}(k+1,l+1)}_{p=q=1+\rho}
\\
&& \hspace{-1.5cm}
=\pmatrix{
(1+\rho)^2 &0 &\cdots &0 \cr
\displaystyle\frac{2\rho}{2!} &(1+\rho)^2 &\cdots &0 \cr
\vdots &\vdots &\ddots &\vdots \cr
\displaystyle\frac{2\rho}{(2N-2)!} &\displaystyle\frac{2\rho}{(2N-4)!}
&\cdots &(1+\rho)^2}
\\
&& \hspace{-1.5cm}
\times\pmatrix{
m_{11}^{(n)}(k,l) &m_{13}^{(n)}(k,l)
&\cdots &m_{1,2N-1}^{(n)}(k,l) \cr
m_{31}^{(n)}(k,l) &m_{33}^{(n)}(k,l)
&\cdots &m_{3,2N-1}^{(n)}(k,l) \cr
\vdots &\vdots &\ddots &\vdots \cr
m_{2N-1,1}^{(n)}(k,l) &m_{2N-1,3}^{(n)}(k,l)
&\cdots &m_{2N-1,2N-1}^{(n)}(k,l)}_{p=q=1+\rho}
\\
&& \hspace{-1.5cm}
\times\pmatrix{
(1+\rho)^2 &\displaystyle\frac{2\rho}{2!}
&\cdots &\displaystyle\frac{2\rho}{(2N-2)!} \cr
0 &(1+\rho)^2 &\cdots &\displaystyle\frac{2\rho}{(2N-4)!} \cr
\vdots &\vdots &\ddots &\vdots \cr
0 &0 &\cdots &(1+\rho)^2}.
\end{eqnarray*}
By taking determinant of both sides, the reduction condition (\ref{red})
is proved. \hfill\fbox{}

\noindent{\bf Proof of Theorem 1} \hspace{0.2cm}  Since the $\tau$
function (\ref{e:taulemma3}) in Lemma 3 satisfies both the bilinear
equations (\ref{bilin}) and the reduction condition (\ref{red}), it
satisfies all the following bilinear equations,
\begin{eqnarray*}
&& \hspace{-1.5cm} (D_x+1)\tau_n(k,l)\cdot\tau_n(k+1,l)=\tau_{n+1}(k,l)\tau_{n-1}(k+1,l),
\\
&& \hspace{-1.5cm} (D_x+1)\tau_n(k,l+1)\cdot\tau_n(k,l)=\tau_{n+1}(k,l+1)\tau_{n-1}(k,l),
\\
&& \hspace{-1.5cm} (D_y-1)\tau_n(k,l+1)\cdot\tau_n(k,l)=-\tau_{n-1}(k,l+1)\tau_{n+1}(k,l),
\\
&& \hspace{-1.5cm} (D_y-1)\tau_n(k,l)\cdot\tau_n(k+1,l)=-\tau_{n-1}(k,l)\tau_{n+1}(k+1,l),
\\
&& \hspace{-1.5cm} \tau_{n+1}(k,l)\tau_{n-1}(k,l)-(1-\rho^2)\tau_n(k,l)\tau_n(k,l)
=\frac{\rho^2}{(1+\rho)^{4N}}\tau_n(k+1,l)\tau_n(k,l+1).
\end{eqnarray*}
We now substitute $x=ict/(1-\rho^2)$ and $y=-idt/(1-\rho^2)$, where
$c$ and $d$ are complex constants. Then the time derivative becomes
$i(1-\rho^2)\partial_t=-c\partial_x+d\partial_y$, and we obtain
\begin{eqnarray*}
&& \hspace{-1cm} [i(1-\rho^2)D_t+c+d]\tau_n(k+1,l)\cdot\tau_n(k,l) \\ && \hspace{-1cm}
=c\tau_{n-1}(k+1,l)\tau_{n+1}(k,l)+d\tau_{n+1}(k+1,l)\tau_{n-1}(k,l),
\\
&& \hspace{-1cm} [-i(1-\rho^2)D_t+c+d]\tau_n(k,l+1)\cdot\tau_n(k,l)  \\ && \hspace{-1cm}
=c\tau_{n+1}(k,l+1)\tau_{n-1}(k,l)+d\tau_{n-1}(k,l+1)\tau_{n+1}(k,l),
\\
&& \hspace{-1cm} \tau_{n+1}(k,l)\tau_{n-1}(k,l)-(1-\rho^2)\tau_n(k,l)\tau_n(k,l)
=\frac{\rho^2}{(1+\rho)^{4N}}\tau_n(k+1,l)\tau_n(k,l+1).
\end{eqnarray*}
The determinant solution (\ref{e:taulemma3}) is now written as
$$
\tau_n(k,l)=\left.\det_{1\le i,j\le N}\left(A_{2i-1}B_{2j-1}m^{(n)}(k,l)
\right)\right|_{p=q=1+\rho},
$$
where operators $A_i$, $B_j$ are defined in Eqs.
(\ref{d:Ai2})-(\ref{d:Bj2}), and
$$
m^{(n)}(k,l)=\frac{1}{pq-1+\rho^2}(pq)^n
\left(\frac{1-\rho^2-q}{1-1/p}\right)^k
\left(\frac{1-\rho^2-p}{1-1/q}\right)^l
e^{i\left(\frac{1}{pq}-\frac{1}{1-\rho^2}\right)\left(qd-pc\right)t}.
$$
By taking $b_{\mu}=\bar a_{\mu}$ and $d=\bar c$, the conjugacy
condition
$$
\tau_n(l,k)=\overline{\tau_n(k,l)}
$$
is then satisfied. We now define
\begin{equation}  \label{def:fngn}
f_n=\tau_n(0,0), \hspace{0.3cm} g_n=\tau_n(1,0)/(1+\rho)^{2N},
\end{equation}
then $f_n$ is real, $\tau_n(0,1)/(1+\rho)^{2N}=\bar g_n$, and the
above bilinear equations yield
\begin{eqnarray*}
&&[i(1-\rho^2)D_t+c+\bar c]g_n\cdot f_n
=cg_{n-1}f_{n+1}+\bar cg_{n+1}f_{n-1},
\\
&&f_{n+1}f_{n-1}-(1-\rho^2)f_nf_n=\rho^2g_n\bar g_n.
\end{eqnarray*}
Finally we set $c=e^{-i\theta}$, where $\theta$ is a real constant.
Then through the variable transformation,
$$
u_n=\frac{\rho}{\sqrt{1-\rho^2}}\frac{g_n}{f_n}e^{i(\theta n-\omega t)},
$$
where
$\omega=(e^{i\theta}+e^{-i\theta})/(1-\rho^2)=2\cos\theta/(1-\rho^2)$,
the above bilinear equations are transformed to
$$
i\frac{d}{dt}u_n=(1+\sigma |u_n|^2)(u_{n+1}+u_{n-1}),
$$
where $\sigma=\mbox{sgn}(1-\rho^2)$. Thus when $|\rho|<1$, the
transformed equation is the focusing AL equation (\ref{e:ALf}),
while when $|\rho|>1$, the transformed equation is the defocusing AL
equation (\ref{e:ALd}). Then rogue wave solutions (\ref{f:general})
for the focusing and defocusing AL equations are established [it is
easy to see that functions $f_n, g_n$ defined in Eq.
(\ref{def:fngn}) are identical to those given in Theorem 1]. The
selection of parameters (\ref{e:aeven}) can be proved in the same
way as that for rogue waves of the NLS equation in \cite{OY} and is
thus not repeated here.~\hfill\fbox{}

\noindent{\bf Proof of Theorem 2}  \hspace{0.2cm} By the
reparametrization $p=1+\rho P$ and $q=1+\rho Q$, the matrix element
$m_{ij}^{(n)}(k,l)=A_iB_jm^{(n)}(k,l)$ in Lemma 2 is given by
$$
m^{(n)}(k,l)=\frac{(-1)^{k+l}\rho^{k+l-1}}{P+Q+\rho(1+PQ)}
(1+\rho P)^{n+k}(1+\rho Q)^{n+l}
\left(\frac{1+Q/\rho}{P}\right)^k\left(\frac{1+P/\rho}{Q}\right)^l
e^{\xi+\eta},
$$
$$
\xi+\eta=\left(1+\rho P-\frac{1-\rho^2}{1+\rho Q}\right)x
+\left(1+\rho Q-\frac{1-\rho^2}{1+\rho P}\right)y,
$$
and
$$
A_i=\sum_{\nu=0}^i\frac{a_{\nu}}{(i-\nu)!}(P\partial_P)^{i-\nu},
\quad
B_j=\sum_{\mu=0}^j\frac{b_{\mu}}{(j-\mu)!}(Q\partial_Q)^{j-\mu}.
$$
Let us consider the following generator ${\cal G}$ of differential
operators $(P\partial_P)^{\alpha}(Q\partial_Q)^{\beta}$,
\[
{\cal G}=\sum_{\alpha=0}^{\infty}\sum_{\beta=0}^{\infty}
\frac{\kappa^\alpha}{\alpha!}\frac{\lambda^\beta}{\beta!}
(P\partial_P)^{\alpha}(Q\partial_Q)^{\beta}
=\exp(\kappa P\partial_P+\lambda Q\partial_Q).
\]
Recalling the identity
$$
{\cal G} F(P, Q)=F(e^k P, e^l Q)
$$
for any function $F$, applying ${\cal G}$ to the above $m^{(n)}$ and
taking $P=Q=1$ (i.e. $p=q=1+\rho$), we have
\begin{eqnarray} \label{gm}
&& \hspace{-2cm} \left.{\cal G}m^{(n)}(k,l)\right|_{P=Q=1}  \nonumber \\ &&   \hspace{-2cm}
=\frac{(-1)^{k+l}\rho^{k+l-1}}
{e^\kappa+e^\lambda+\rho(1+e^{\kappa+\lambda})}
(1+\rho e^\kappa)^{n+k}(1+\rho e^\lambda)^{n+l}
\left(\frac{1+e^\lambda/\rho}{e^\kappa}\right)^k\left(\frac{1+e^\kappa/\rho}{e^\lambda}\right)^l
e^{\tilde\xi+\tilde\eta}
\nonumber\\
&& \hspace{-2cm}
=\frac{(-1)^{k+l}(1+\rho)^{2(n+k+l)-1}/2\rho}
{\displaystyle 1-\frac{1-\rho}{1+\rho}\hspace{0.1cm}
\frac{1-e^\kappa}{1+e^\kappa} \hspace{0.1cm} \frac{1-e^\lambda}{1+e^\lambda}}
\exp\left[(n+k)\ln\frac{1+\rho e^\kappa}{1+\rho}
+(n+l)\ln\frac{1+\rho e^\lambda}{1+\rho}\right.
\nonumber\\
&& \hspace{-2cm} \quad
\left.+k\ln\frac{1+e^\lambda/\rho}{1+1/\rho}
+l\ln\frac{1+e^\kappa/\rho}{1+1/\rho}-k\kappa-l\lambda
-\ln\frac{1+e^\kappa}{2}-\ln\frac{1+e^\lambda}{2}
+\tilde\xi+\tilde\eta\right],
\end{eqnarray}
where
\[
\tilde\xi+\tilde\eta
=\left(1+\rho e^\kappa-\frac{1-\rho^2}{1+\rho e^\lambda}\right)x
+\left(1+\rho e^\lambda-\frac{1-\rho^2}{1+\rho e^\kappa}\right)y.
\]
Differentiating the first expansion in (\ref{rs}) with respect to
$\lambda$, we get
\[
\frac{\rho e^\lambda}{1+\rho e^\lambda}=\frac{\rho}{1+\rho}
+\sum_{\nu=1}^\infty(\nu+1)r_{\nu+1}(\rho)\lambda^\nu,
\]
thus $\tilde\xi+\tilde\eta$ can be written as
\begin{eqnarray*}
&&
\tilde\xi+\tilde\eta=2\rho(x+y)
+\left(\rho(e^\kappa-1)+(1-\rho^2)
\sum_{\nu=1}^\infty(\nu+1)r_{\nu+1}(\rho)\lambda^\nu\right)x  \\ &&  \hspace{1.3cm}
+\left(\rho(e^\lambda-1)+(1-\rho^2)
\sum_{\nu=1}^\infty(\nu+1)r_{\nu+1}(\rho)\kappa^\nu\right)y.
\end{eqnarray*}
Moreover since we have the formal expansion
$$
\left[\displaystyle 1-\frac{1-\rho}{1+\rho}\hspace{0.1cm}
\frac{1-e^\kappa}{1+e^\kappa}\hspace{0.1cm} \frac{1-e^\lambda}{1+e^\lambda}\right]^{-1}
=\sum_{\mu=0}^\infty
\left[\frac{1-\rho}{1+\rho} \hspace{0.1cm} \frac{\kappa\lambda}{4}
\exp\left(\ln(\frac{2}{\kappa}\tanh\frac{\kappa}{2})
+\ln(\frac{2}{\lambda}\tanh\frac{\lambda}{2})\right)\right]^\mu,
$$
Eq. (\ref{gm}) can be rewritten as
\begin{eqnarray*}
&&\frac{(-1)^{k+l}2\rho e^{-2\rho(x+y)}}{(1+\rho)^{2(n+k+l)-1}}
\left.{\cal G}m^{(n)}(k,l)\right|_{P=Q=1}  \\ &&
=\sum_{\mu=0}^\infty
\left(\frac{1-\rho}{1+\rho}\hspace{0.1cm} \frac{\kappa\lambda}{4}\right)^\mu
\exp\left(\sum_{\nu=1}^\infty(x_\nu+\mu s_\nu)\kappa^\nu
+\sum_{\nu=1}^\infty(y_\nu+\mu s_\nu)\lambda^\nu\right)
\\
&&
=\sum_{\mu=0}^\infty
\left(\frac{1-\rho}{1+\rho} \hspace{0.1cm} \frac{\kappa\lambda}{4}\right)^\mu
\sum_{\alpha=0}^\infty
S_\alpha(\mbox{\boldmath $x$}+\mu\mbox{\boldmath $s$})\kappa^\alpha
\sum_{\beta=0}^\infty
S_\beta(\mbox{\boldmath $y$}+\mu\mbox{\boldmath $s$})\lambda^\beta,
\end{eqnarray*}
where $x_\nu$ and $y_\nu$ are as defined in Theorem 2. By taking the
coefficient of $\kappa^\alpha\lambda^\beta$, we obtain
\begin{eqnarray*}
&&
\frac{(-1)^{k+l}2\rho e^{-2\rho(x+y)}}{(1+\rho)^{2(n+k+l)-1}}
\left.\frac{(P\partial_P)^{\alpha}}{\alpha!}
\frac{(Q\partial_Q)^{\beta}}{\beta!}m^{(n)}(k,l)\right|_{P=Q=1}  \\ &&
=\sum_{\mu=0}^{\min(\alpha,\beta)}
\frac{1}{4^\mu}\left(\frac{1-\rho}{1+\rho}\right)^\mu
S_{\alpha-\mu}(\mbox{\boldmath $x$}+\mu\mbox{\boldmath $s$})
S_{\beta-\mu}(\mbox{\boldmath $y$}+\mu\mbox{\boldmath $s$}).
\end{eqnarray*}
Therefore the matrix element $m_{ij}^{(n)}(k,l)$ in Lemma 2 with
$p=q=1+\rho$ is explicitly expressed in the polynomial form,
\begin{eqnarray*}
&& \hspace{-1cm}  \left.\frac{(-1)^{k+l}2\rho e^{-2\rho(x+y)}}{(1+\rho)^{2(n+k+l)-1}}
m_{ij}^{(n)}(k,l)\right|_{p=q=1+\rho}   \nonumber \\ && \hspace{-1cm}
=\sum_{\alpha=0}^i\sum_{\beta=0}^j
\sum_{\mu=0}^{\min(i-\alpha,j-\beta)}
\frac{a_\alpha b_\beta}{4^\mu}\left(\frac{1-\rho}{1+\rho}\right)^\mu
S_{i-\alpha-\mu}(\mbox{\boldmath $x$}+\mu\mbox{\boldmath $s$})
S_{j-\beta-\mu}(\mbox{\boldmath $y$}+\mu\mbox{\boldmath $s$}).
\end{eqnarray*}
By taking $x=ite^{-i\theta}/(1-\rho^2)$,
$y=-ite^{i\theta}/(1-\rho^2)$ and $b_\beta=\bar a_\beta$, the matrix
element $m_{ij}^{(n)}(k,l)$ from the above equation is equal to
$\widetilde m_{ij}^{(n)}(k,l)$ in Theorem 2, multiplied by a factor
which is $(i,j)$-independent and is inversely proportional to
$(-1)^k(1+\rho)^{2k}$. Recalling the definition of functions $f_n,
g_n$ in Eq. (\ref{def:fngn}), $g_n/f_n$ in Theorem 1 is then equal
to $(-1)^N\sigma_n(1,0)/\sigma_n(0,0)$, thus the algebraic
expression of rogue waves in Eqs.
(\ref{f:theorem2})-(\ref{def:sigman}) of Theorem 2 is proved.

The alternative expression (\ref{sig}) for $\sigma_n(k,l)$ in
Theorem 2 can be derived directly from the original expression
(\ref{def:sigman}) through a similar determinant calculus in
\cite{OY}. We rewrite the $N\times N$ determinant $\sigma_n(k,l)$ in
(\ref{def:sigman}) into a $3N\times 3N$ determinant form, then apply
the Laplace expansion, which leads to the expression (\ref{sig})
(see \cite{OY} for details).  Thus, Theorem 2 is proved.
\hfill\fbox{}

\noindent{\bf Proof of Theorem 3}  \hspace{0.2cm} In the polynomial
solution (\ref{sig}) in $n$ and $t$, the leading term comes from the
one with $\nu_1=0$, $\nu_2=1$, $\ldots$, $\nu_N=N-1$, i.e.,
$$
[(1-\rho)/(1+\rho)]^{N(N-1)/2}
\det_{1\le i,j\le N}\left(\Phi_{2i-1,j-1}^{(n)}(k,l)\right)
\det_{1\le i,j\le N}\left(\Psi_{2i-1,j-1}^{(n)}(k,l)\right).
$$
Notice that the highest-degree terms in $\Phi_{i\nu}^{(n)}(k,l)$ and
$\Psi_{j\nu}^{(n)}(k,l)$ are $a_0x_1^{i-\nu}/(i-\nu)!2^\nu$ and
$\bar a_0y_1^{j-\nu}/(j-\nu)!2^\nu$ respectively. In addition, the
leading term of $n$, $t$ in the product $x_1y_1$ is given by
$|r_1(\rho)n+i(e^{-i\theta}\rho/(1-\rho^2)-e^{i\theta}2r_2(\rho))t|^2
=r_1(\rho)^2|n+i(e^{-i\theta}/(1-\rho)-e^{i\theta}/(1+\rho))t|^2$.
Consequently the leading term of $n$, $t$ in the polynomial
$\sigma_n(k,l)$ is proportional to
$$
\left|n+i\left(\frac{e^{-i\theta}}{1-\rho}
-\frac{e^{i\theta}}{1+\rho}\right)t\right|^{N(N+1)}
=\left[\left(n+\frac{2t\sin\theta}{1-\rho^2}\right)^2
+\left(\frac{2t\rho\cos\theta}{1-\rho^2}\right)^2\right]^{N(N+1)/2}.
$$
If $\cos\theta \ne 0$, then this term is dominant as $n^2+t^2$ goes
to infinity in any direction on the $(n,t)$-plane. The coefficient
of this dominant term does not vanish and is independent of $k$ and
$l$ by direct calculation. Thus if $\cos\theta \ne 0$,
$\sigma_n(1,0)/\sigma_n(0,0)$ approaches 1 when the space-time point
$(n,t)$ goes to infinity, for example, when $t$ goes to infinity
(for each fixed $n$) or $n$ goes to infinity (for each fixed $t$).
In addition, if $\cos\theta \ne 0$, then
$\sigma_n(1,0)/\sigma_n(0,0)$ approaches 1 uniformly in $n$ as $|t|$
goes to infinity.  Hence the boundary condition (\ref{e:bc}) in
Theorem 3 is proved. \hfill\fbox{}

\noindent{\bf Proof of Theorem 4}  \hspace{0.2cm} From Theorem 2, we
see that $\Psi_{i\nu}^{(n)}(k,l)$ is the complex conjugate of
$\Phi_{i\nu}^{(n)}(l,k)$. Then from the expression of
$\sigma_n(k,l)$ in Eq. (\ref{sig}), we have
$$
\sigma_n(0,0)=\sum_{\nu_1=0}^1\sum_{\nu_2=\nu_1+1}^3
\cdots\sum_{\nu_N=\nu_{N-1}+1}^{2N-1}
\left(\frac{1-\rho}{1+\rho}\right)^{\nu_1+\nu_2+\cdots+\nu_N}
\left|\det_{1\le i,j\le N}\left(\Phi_{2i-1,\nu_j}^{(n)}(0,0)\right)\right|^2.
$$
Clearly $\sigma_n(0,0)\ge0$ if $-1<\rho<1$. Furthermore the term for
$\nu_1=1$, $\nu_2=3$, $\ldots$, $\nu_N=2N-1$ is not zero because
$\Phi_{ii}^{(n)}(k,l)=a_0/2^i$ and $\Phi_{i\nu}^{(n)}(k,l)=0$
$(i<\nu)$. Therefore $\sigma_n(0,0)$ is strictly positive for
$-1<\rho<1$. Consequently, rogue waves for the focusing AL equation
in Theorems 1 and 2 are always non-singular, which completes the
proof.~\hfill\fbox{}

Of course, this non-singularity of solutions does not hold for the
defocusing AL equation (where $|\rho|>1$), as has been seen in
Sec.~3.

\section{Summary}

In this paper, general $N$-th order rogue waves in the focusing and
defocusing Ablowitz-Ladik equations were derived by the bilinear
method. These solutions were given by determinants, and they contain
$2N+1$ non-reducible free real parameters. In the focusing case, we
showed that rogue waves are always bounded. In addition, they can
reach much higher peak amplitudes than their continuous counterparts
in the NLS equation. Furthermore, higher-order rogue waves can
exhibit triangular and circular patterns with different individual
peaks. In the defocusing case, we showed that rogue waves still
appear, which is surprising. In this case, we found that rogue waves
of any order can blow up to infinity in finite time, even though
non-blowup rogue waves can also exist.

It is noted that solutions in the AL equations are closely related
to those in the discrete Hirota equation \cite{Ankiewicz2010}
\begin{equation} \label{e:dHirota}
i\frac{d}{dt}v_n=(1\pm |v_n|^2)[(a+ib)v_{n+1}+(a-ib)v_{n-1}],
\end{equation}
where $a$ and $b$ are any real constants. Since we have constructed
rogue-wave solutions of the AL equations for arbitrary carrier wave
frequency $\theta$, it is straightforward to derive rogue-wave
solutions in the discrete Hirota equation as well. By writing
$a+ib=Re^{i\Theta}$ with real constants $R$ and $\Theta$, we see
that
$$
v_n(t)=e^{-in\Theta}u_n(Rt)
$$
is a solution of Eq. (\ref{e:dHirota}) when $u_n(t)$ is a solution
of the AL equation (\ref{e:ALf}) or (\ref{e:ALd}). Thus, rogue-wave
solutions for the discrete Hirota equation can be obtained directly
from Theorem 1 or 2. For example, from Theorem 2 we get the
following theorem. Here for simplifying the expression, the sign
$(-1)^N$ in Theorem 2 is dropped and notations $\phi=\theta-\Theta$,
$\Omega=R\omega$ are used.

\noindent\textbf{Theorem 5}\hspace{0.2cm} General $N$-th order rogue
waves for the discrete Hirota equation (\ref{e:dHirota}) are given
by
$$
v_n(t)=\frac{\rho}{\sqrt{1-\rho^2}}\:
\frac{\sigma_n(1,0)}{\sigma_n(0,0)} \hspace{0.15cm}  e^{i(\phi n-\Omega t)},
$$
where $\rho$, $\phi$ are free real constants,
$\Omega=[(a+ib)e^{i\phi}+(a-ib)e^{-i\phi}]/(1-\rho^2)$, and
$\sigma_n(k,l)$ is defined in Theorem 2 with $x$ and $y$ replaced by
$it(a-ib)e^{-i\phi}/(1-\rho^2)$ and $-it(a+ib)e^{i\phi}/(1-\rho^2)$
respectively.

\section*{Acknowledgment}
The work of Y.O. was supported in part by JSPS grant-in-aid for
Scientific Research (B-24340029, S-24224001) and for Challenging
Exploratory Research (22656026), and the work of J.Y. was supported
in part by the Air Force Office of Scientific Research (grant USAF
9550-12-1-0244) and the National Science Foundation (grant
DMS-1311730).

\end{document}